\documentclass[12pt]{article}
\usepackage[dvips]{graphicx}
\usepackage{pdproc}
\makeatletter 
\def\@cite#1{[#1]} 
\makeatother    
\newcommand {\nn}{\nonumber \\}

\newcommand {\tr}{{\rm tr\,}}

\newcommand {\pl}{\partial}

\newcommand {\vp}{\varphi}

\newcommand {\al}{\alpha}
\newcommand {\be}{\beta}
\newcommand {\ga}{\gamma}

\newcommand {\la}{\lambda}

\newcommand {\Th}{\Theta}

\newcommand {\ep}{\epsilon}

\newcommand {\na}{\nabla}
\newcommand {\del}  {\delta}

\newcommand {\ab}   {{\alpha\beta}}

\newcommand {\half}{ {\frac{1}{2}} }

\newcommand {\sqtwo} {\sqrt{2}}

\newcommand {\Lcal}{{\cal L}}

\newcommand {\Dcal}{{\cal D}}
\newcommand {\Ncal}{{\cal N}}

\def\overleftarrow#1{\vbox{\ialign{##\crcr
 $\leftarrow$\crcr\noalign{\kern-1pt\nointerlineskip}
 $\hfil\displaystyle{#1}\hfil$\crcr}}}

\newcommand {\eptil}{{\tilde \epsilon}}

\newcommand {\abar}{{\bar a}}

\newcommand {\cbar}{{\bar c}}

\newcommand {\Bbar}{{\bar B}}

\newcommand {\labar}{{\bar \lambda}}
\newcommand {\psibar}{{\bar \psi}}
\newcommand {\vpbar}{{\bar \varphi}}

\newcommand {\sibar}{{\bar \sigma}}

\newcommand {\alp}{{\alpha'}}
\newcommand {\bep}{{\beta'}}
\newcommand {\gap}{{\gamma'}}
\newcommand  {\xf}{{x^5}}

\newcommand {\change} {\leftrightarrow}
\newcommand {\ra} {\rightarrow}

\newcommand {\com}  {{\quad ,}}
\newcommand {\q}    {\quad}

\newcommand {\nl}    {\newline}

\newcommand {\NP}   {Nucl.Phys.}

\begin{document}

\title{
 Family of Singular Solutions in a SUSY 
 Bulk-Boundary System
\footnote{
Based on the work with AKIHIRO MURAYAMA[1]
}}

\author{ SHOICHI ICHINOSE}

\address{ 
School of Food and Nutritional Sciences, University of Shizuoka\\
Yada 52-1, Shizuoka 422-8526, Japan
%%%%% You may comment out the e-mail address line below.  
\\ {\rm E-mail: ichinose@u-shizuoka-ken.ac.jp}}

\abstract{
A set of classical solutions of a singular type
is found in a 5D SUSY bulk-boundary system.
The "parallel" configuration, where the whole components
of fields or branes are parallel in the iso-space, 
naturally appears. 
It has three {\it free} parameters related to
the {\it scale freedom} in the choice of the 
brane-matter sources and the {\it "free" wave} property
of the {\it extra component} of the bulk-vector field.
The solutions describe brane, anti-brane and brane-anti-brane
configurations depending on the parameter choice.
Some solutions describe the localization behaviour
even after the non-compact limit of the extra space. 
Stableness is assured. 
Their meaning in the brane world physics is
examined.
}

\normalsize\baselineskip=15pt
\vspace{3mm}
%%%%%%%%%%%%%%%%%%%%%%%%%%%%%%%%%%%%%%%%%%%%%%%%%%%%%%%%%%%%%%%%%%%%%%
%%%%%%%%%%% Mirabelli-Peskin Model                   %%%%%%%%%%%%%%%%%
%%%%%%%%%%%%%%%%%%%%%%%%%%%%%%%%%%%%%%%%%%%%%%%%%%%%%%%%%%%%%%%%%%%%%%

{\bf 1}\ {\it Mirabelli-Peskin Model}\q
Let us consider the 5 dimensional flat space-time with the signature
(-1,1,1,1,1).
The space of the fifth
component is taken to be $S_1$, 
with the periodicity $2l$, and has the 
$Z_2$-orbifold condition($\xf\change -\xf$).
We take a 
5D bulk theory $\Lcal_{bulk}$ which is
coupled with a 4D matter theory $\Lcal_{bnd}$ on a "wall" at $\xf=0$
and with $\Lcal'_{bnd}$ on the other "wall" at $\xf=l$:\ 
%The boundary Lagragians are, in the bulk action,  described by
% the delta-functions along the extra axis $x^5$.
%*** mp1 %%%%%%%%%%%%%%%%
%\begin{eqnarray}
$S=\int_{-l}^{l}dx^5\int d^4x\{\Lcal_{blk}+\del(x^5)\Lcal_{bnd}
+\del(x^5-l){\Lcal'}_{bnd}
\}$. 
%\pr\label{mp1}
%\end{eqnarray}
%%%%%%%%%%%%%%%%%%%%%%%%%%%%%
%We consider both bulk and boundary quantum effects.

We take the Mirabelli-Peskin model[2] as an example. 
The bulk dynamics is given by the 5D super YM theory(
a vector field $A^M\ (M=0,1,2,3,5)$, 
a scalar field $\Phi$, 
a doublet of symplectic Majorana fields $\la^i\ (i=1,2)$, 
and a triplet of auxiliary scalar fields $X^a\ (a=1,2,3)$):
%*** mp2 %%%%%%%%%%%%%%%%
%\begin{eqnarray}
$\Lcal_{SYM}=-\half\tr {F_{MN}}^2-\tr (\na_M\Phi)^2
-i\tr(\labar_i\ga^M\na_M\la^i)
+\tr (X^a)^2+g\,\tr (\labar_i[\Phi,\la^i])$.    %\nn
%F_{MN}=\pl_MA_N-\pl_NA_M-ig[A_M,A_N]\com\q
%D_M\Phi=\pl_M\Phi-ig[A_M,\Phi]\com\nn
%D_M\la^i=\pl_M\la^i-ig[A_M,\la^i]
%\com
%\label{mp2}
%\end{eqnarray}
%%%%%%%%%%%%%%%%%%%%%%%%%%%%%
%where all bulk fields are the {\it adjoint} representation
%(suffixes: $\al,\be,\cdots $)
%of the gauge group $G$. 
%*** mp4 %%%%%%%%%%%%%%%%
%\begin{eqnarray}
%\del_\xi A^M=i\xibar^i\ga^M\la^i\com\nn
%\del_\xi\Phi=i\xibar^i\la^i\com\nn
%\del_\xi\la^i=(\Si^{MN}F_{MN}-\ga^MD_M\Phi)\xi^i
%-i(X^a\si^a)^{ij}\xi^j\com\nn
%\del_\xi X^a=\xibar^i(\si^a)^{ij}\ga^MD_M\la^j
%-i[\Phi,\xibar^i(\si^a)^{ij}\la^j]\com
%\label{mp4}
%\end{eqnarray}
%%%%%%%%%%%%%%%%%%%%%%%%%%%%%
%where $\Si^{MN}=\fourth [\ga^M,\ga^N]$, and
%the SUSY global parameter $\xi^i$ is the symplectic Majorana
%spinor. 
%As the 5D gauge-fixing term, we take the Feynman
%*** Landau??? cf Barbieri et al PLB82**** 
%gauge:
%*** mp4b %%%%%%%%%%%%%%%%
%\begin{eqnarray}
%\Lcal_{gauge}=-\tr (\pl_MA^M)^2=-\half (\pl_MA^M_{~\al})^2
%\pr
%\label{mp4b}
%\end{eqnarray}
%%%%%%%%%%%%%%%%%%%%%%%%%%%%%
%The corresponding ghost Lagrangian is given by
%*** mp4c %%%%%%%%%%%%%%%%
%\begin{eqnarray}
%\Lcal_{ghost}=-2\,\tr \pl_M\cbar\cdot \na^M(A)c
%=-2\,\tr\pl_M\cbar\cdot (\pl^Mc+ig[A^M,c])
%\com
%\label{mp4c}
%\end{eqnarray}
%%%%%%%%%%%%%%%%%%%%%%%%%%%%%
%where $c$ and $\cbar$ are the complex ghost fields. 
%We take the following bulk action.
%*** mp4d %%%%%%%%%%%%%%%%
%\begin{eqnarray}
%\Lcal_{blk}=\Lcal_{SYM}+\Lcal_{gauge}+\Lcal_{ghost}
%\pr
%\label{mp4d}
%\end{eqnarray}
%%%%%%%%%%%%%%%%%%%%%%%%%%%%%

It is known that we can consistently project out $\Ncal=1$ SUSY
multiplet %, which has 4 real super charges, 
%*** mp5 %%%%%%%%%%%%%%%%
%\begin{eqnarray}
%Z_2\mbox{ transformation}:\ 
%x^5\ra -x^5\pr
%\label{mp5}
%\end{eqnarray}
%%%%%%%%%%%%%%%%%%%%%%%%%%%%%
by assigning $Z_2$-parity 
to all fields in accordance with the 5D SUSY. 
A consistent choice is given as:\  $P=+1$ for 
$A^m, \la_L, X^3$;  $P=-1$ for 
$A^5, \Phi, \la_R, X^1, X^2$ ($m=0,1,2,3$). 
Then ($A^m,\la_L,X^3-\na_5\Phi$) constitute
 an $\Ncal =1$ vector multiplet. 
Especially $\Dcal\equiv X^3-\na_5\Phi$ plays the role
of {\it D-field} on the wall. 
We introduce one 4D chiral multiplet ($\phi,\psi,F$) on the $\xf=0$ wall
and the other one ($\phi',\psi',F'$) on the $\xf=l$ wall. 
These are the simplest matter candidates and were taken
in the original paper[2]%\cite{MP97}
. 
Using the $\Ncal=1$ SUSY property of the fields 
($A^m,\la_L,X^3-\na_5\Phi$),
we can find the following bulk-boundary coupling on the $\xf=0$ wall.
%*** mp7 %%%%%%%%%%%%%%%%
\begin{eqnarray}
\Lcal_{bnd}=-\na_m\phi^\dag \na^m\phi-\psi^\dag i\sibar^m \na_m\psi+F^\dag F%\nn
+\sqrt{2}ig(\psibar\labar_L\phi-\phi^\dag\la_L\psi)
+g\phi^\dag \Dcal\phi
+\Lcal_{SupPot}
\ ,\nn
\Lcal_{SupPot}=
(\half m_{\alp\bep}\Th_\alp\Th_\bep
+\frac{1}{3!}\la_{\alp\bep\gap}\Th_\alp\Th_\bep\Th_\gap)|_{\theta^2}
+\mbox{h.c.}
\com
\label{mp7}
\end{eqnarray}
%%%%%%%%%%%%%%%%%%%%%%%%%%%%%
where $    %c=-i\si^2,\ 
\na_m\equiv \pl_m+igA_m,\ \Dcal =X^3-\na_5\Phi,\ 
\Th=\phi+\sqtwo\theta\psi+\theta^2 F
$.
We take the {\it fundamental} representation for $\Th=(\phi,\psi,F)$. 
%The quadratic (kinetic) terms of the vector $A^m$, the gaugino spinor $\la_L$
%and the 'auxiliary' field $\Dcal=X^3-\na_5\Phi$ are in the bulk world. 
In the same way, we introduce the coupling between 
the other matter fields
($\phi',\psi',F'$) on the $\xf=l$ wall and the bulk fields:\ 
$\Lcal'_{bnd}=(\phi\ra\phi', \psi\ra\psi', F\ra F' \ in\ \mbox{(\ref{mp7})})$. 
We note the interaction between the bulk fields and the boundary
ones is {\it definitely fixed from SUSY}.

%%%%%%%%%%%%%%%%%%%%%%%%%%%%%%%%%%%%%%%%%%%%%%%%%%%%%%%%%%%%%%%%%%%%%%
%%%%%%%%%%% Vacuum of Mirabelli-Peskin Model         %%%%%%%%%%%%%%%%%
%%%%%%%%%%%%%%%%%%%%%%%%%%%%%%%%%%%%%%%%%%%%%%%%%%%%%%%%%%%%%%%%%%%%%%
{\bf 2}\ {\it Vacuum Structure}\q
%We now examine the vacuum structure. 
Generally the vacuum is
determined by the potential part of scalar fields.
We first reduce the previous system to the part
which involves only scalar fields or the extra
component of the bulk vector. 
%*** ep3 %%%%%%%%%%%%%%%%
\begin{eqnarray}
\Lcal^{red}_{blk}[\Phi,X^3,A_5]
%=\half\pl_M\Phi_\al\pl^M\Phi_\al
%+\half X^3_\al X^3_\al\nn
%+\half\pl_MA_{5\al}\pl^MA_{5\al}  %**** M\ra m???? ****\nn
%-gf_{\ab\ga}\pl_5\Phi_\al\cdot A_{5\be}\Phi_\ga
%-\frac{g^2}{2}f_{\ab\tau}f_{\ga\del\tau}A_{5\al}\Phi_\be A_{5\ga}\Phi_\del\nn
%\mbox{OR}\nn
=\tr \left\{ -\pl_M\Phi\pl^M\Phi+X^3X^3-\pl_MA_5\pl^MA_5
+2g(\pl_5\Phi\times A_5)\Phi\right.\nn
\left. -g^2(A_5\times\Phi)(A_5\times\Phi)
-2\pl_M\cbar\cdot\pl^Mc-2ig\pl_5\cbar\cdot [A^5,c]\right\}
%+\mbox{irrel. terms}
\com
\label{ep3}
\end{eqnarray}
%%%%%%%%%%%%%%%%%%%%%%%%%%%%%
where we have dropped terms of $X^1$ and $X^2$
%$2\tr X^1X^1=X^1_\al X^1_\al, 2\tr X^2X^2=X^2_\al X^2_\al$
 because they decouple from other fields. 
%(Note $\tr (\pl_5\Phi\times A_5)\Phi
%=(1/2)f_{\ab\ga}\pl_5\Phi_\al A_{5\be}\Phi_\ga$). 
The field $c$ is the ghost field.  
%which is introduced in the
%usual procedure of fixing the gauge freedom of $\Lcal_{SYM}$.
While
$\Lcal_{bnd}$, on the $\xf=0$ wall, reduces to
%*** ep4 %%%%%%%%%%%%%%%%
\begin{eqnarray}
%\mbox{a. Chiral matter}\nn
\Lcal^{red}_{bnd}[\phi,\phi^\dag,X^3-\na_5\Phi]=-\pl_m\phi^\dag\pl^m\phi
+g(X^3_\al-\na_5\Phi_\al)\phi^\dag_\bep (T^\al)_{\bep\gap}\phi_\gap
+F^\dag F
\nn
+
\left\{
\frac{m_{\alp\bep}}{2}(\phi_\alp F_\bep+F_\alp\phi_\bep)
+\frac{\la_{\alp\bep\gap}}{3!}(\phi_\alp\phi_\bep F_\gap
+\phi_\alp F_\bep\phi_\gap+F_\alp\phi_\bep\phi_\gap)+\mbox{h.c.}
\right\}
%\mbox{b. Non-chiral matter}\nn
%\Lcal^{red}_{bnd(b)}[\phi_S,\phi_S^\dag,\phi_R,\phi_R^\dag,X^3-\pl_5\Phi]=
%\pl_m\phi_S^\dag\pl^m\phi_S+\pl_m\phi_R^\dag\pl^m\phi_R\nn
%+g(X^3_\al-\pl_5\Phi_\al)(T^\al)_{\bep\gap}
%(\phi^\dag_{S\bep} \phi_{S\gap}-\phi^\dag_{R\bep} \phi_{R\gap})
%+\mbox{irrl. terms}
\ .
\label{ep4}
\end{eqnarray}
%%%%%%%%%%%%%%%%%%%%%%%%%%%%%
%$\alp, \bep\cdots$ are the suffixes of the fundamental representation.
In the same way, we obtain 
${\Lcal^{red}_{bnd}}'[\phi',{\phi'}^\dag,X^3-\na_5\Phi]$
on the $\xf=l$ wall by replacing, in (\ref{ep4}), 
$\phi$ and $\phi^\dag$ by $\phi'$ and $\phi'^\dag$, respectively.

The vacuum is usually obtained by the
{\it constant} solution of the scalar-part field equation. 
In higher dimensional models, however, 
extra-coordinate(s) can be regarded as parameter(s)
which should be separately treated 
from the 4D space-time coordinates. 
In this standpoint, it is the more general treatment of
the vacuum that we allow the {\it $\xf$-dependence} on 
the bulk-part of the solution. 
We generally call the classical solutions
($\vp, \chi^3, a_5; \eta, \eta', f, f'$) 
the {\it background fields}. 
\footnote{
In the background field treatment[3]%\cite{IM0312}
 we 
expand all fields around the background fields:\ 
$\vp+\Phi, \chi^3+X^3, a_5+A_5, \eta+\phi, \eta'+\phi', 
f+F, f'+F'$
}
They satisfy 
the {\it field equations} derived from (\ref{ep3}) and (\ref{ep4})\ 
({\it on-shell} condition). 
Assuming 
$\vp=\vp(\xf), \chi^3=\chi^3(\xf), 
a_5=a_5(\xf), \eta=\mbox{const}, \eta'=\mbox{const},
f=\mbox{const}, f'=\mbox{const}$, 
the field equation of 
$
\Lcal^{red}_{blk}+\del(x^5)\Lcal^{red}_{bnd}
+\del(x^5-l){\Lcal^{red}_{bnd}}'
$
are given by, for the bulk-fields variation, 
%*** vac1a,b,c %%%%%%%%%%%%%%%%
\begin{eqnarray}
\del\Phi_\al\q;\ 
-\pl_5Z_\al-g(Z\times a_5)_\al=0\ ,\q 
\del A_{5\al}\q;\ 
{\pl_5}^2a_{5\al}-g(\vp\times Z)_\al=0\ ,
\nn
\del X^3_\al\q;\ 
\chi^3_\al+g(\del(\xf)\eta^\dag T^\al \eta+\del(\xf-l)\eta'^\dag T^\al \eta')
=0\ ,
\com
\label{vac1c}
\end{eqnarray}
%%%%%%%%%%%%%%%%%%%%%%%%%%%%% 
where 
$Z_\al\equiv -g(\del(\xf)\eta^\dag T^\al\eta+\del(\xf-l)\eta'^\dag T^\al\eta')
-\pl_5\vp_\al+gf_{\ab\ga}a_{5\be}\vp_\ga$. 
The field equations for 
the boundary-fields part are given by the variations
$\del\phi^\dag_\alp\q (\del\phi'^\dag_\alp)$ and 
$\del F^\dag_\alp\q (\del F'^\dag_\alp)$:
%*** vac2a,b %%%%%%%%%%%%%%%%
\begin{eqnarray}
{d_\be}|_{\xf=0}\times (T^\be\eta)_\alp
+m_{\alp\bep}f^\dag_\bep+\half\la_{\alp\bep\gap}\eta^\dag_\bep f^\dag_\gap
=0\ ,\ 
(\eta\ra\eta', f\ra f'\ \mbox{in the left equation})\ ,\ \label{vac2a}
\nn
f_\alp+m_{\alp\bep}\eta^\dag_\bep+\half\la_{\alp\bep\gap}
\eta^\dag_\bep\eta^\dag_\gap=0\ ,\ 
(\eta\ra\eta', f\ra f'\ \mbox{in the left equation})
\ ,\ 
\label{vac2b}
\end{eqnarray}
%%%%%%%%%%%%%%%%%%%%%%%%%%%%% 
where $d_\al=(\chi^3-\pl_5\vp+ga_5\times\vp)_\al$ 
is the background D-field. 
From the equation (\ref{vac1c}), we obtain
%*** vac3 %%%%%%%%%%%%%%%%
%\begin{eqnarray}
$\chi^3_\al=-g(\del(\xf)\eta^\dag T^\al \eta+\del(\xf-l)\eta'^\dag T^\al \eta')$,
%\pr
%\label{vac3}
%\end{eqnarray}
%%%%%%%%%%%%%%%%%%%%%%%%%%%%% 
and we know
%*** vac4 %%%%%%%%%%%%%%%%
%\begin{eqnarray}
$Z_\al=d_\al$
%\pr
%\label{vac4}
%\end{eqnarray}
%%%%%%%%%%%%%%%%%%%%%%%%%%%%% 

Before presenting the solution,
we note a simple structure involved in them. Under
the "{\it parallel}" circumstance, 
$a_{5\al}\propto\vp_\al\propto\eta^\dag T^\al\eta
\propto\eta'^\dag T^\al\eta'$, the equations
for $\del\Phi_\al$ 
and $\del A_{5\al}$ 
are
%*** vac4b %%%%%%%%%%%%%%%%
%\begin{eqnarray}
${\pl_5}^2\vp_\al=
-g\pl_5(\del(\xf)\eta^\dag T^\al\eta
+\del(\xf-l)\eta'^\dag T^\al\eta')$
and
${\pl_5}^2a_{5\al}=0$. 
%\pr
%\label{vac4b}
%\end{eqnarray}
%%%%%%%%%%%%%%%%%%%%%%%%%%%%% 
The first one is a static wave equation with "source" fields
located at $\xf=0$ and $l$. 
It is easily integrated once. 
%*** vac4c %%%%%%%%%%%%%%%%
%\begin{eqnarray}
${\pl_5}\vp_\al=
-g(\del(\xf)\eta^\dag T^\al\eta
+\del(\xf-l)\eta'^\dag T^\al\eta')+\mbox{const}$.\ 
%\pr
%\label{vac4c}
%\end{eqnarray}
%%%%%%%%%%%%%%%%%%%%%%%%%%%%% 
This result was used in ref.[2]%\cite{MP97}
. 
The equation of $a_{5\al}$ is  a (static)
"free" wave equation (no source fields). $a_{5\al}$
do {\it not} receive, in the "parallel" environment, any
effect from the boundary sources $\eta, \eta'$.
This characteristically shows the difference between
the bulk scalar $\Phi_\al$ and the extra component
of the bulk vector $A_{5\al}$ in the vacuum
configuration. 

We first solve (\ref{vac1c}) with respect to
$a_{5\al}$ and $\vp_\al$. They also give the solutions
for $\chi^3_\al$ and $d_\al=Z_\al$. Using these results
we solve (\ref{vac2a}) with respect to
$\eta, \eta', f$ and $f'$ for given values of
$m_{\alp\bep}$ and $\la_{\alp\bep\gap}$. 
Here we seek a natural solution by requiring that 
 $d_\al(=Z_\al)$ is {\it independent of $\xf$}. 
%*** vac5 %%%%%%%%%%%%%%%%
%\begin{eqnarray}
%Z_\al=d_\al=-g(\del(\xf)\eta^\dag T^\al\eta+\del(\xf-l)\eta'^\dag T^\al\eta')
%-\pl_5\vp_\al+gf_{\ab\ga}a_{5\be}\vp_\ga=\nn
%\mbox{independent of }\xf\mbox{ (const)}
%\pr
%\label{vac5}
%\end{eqnarray}
%%%%%%%%%%%%%%%%%%%%%%%%%%%%% 
Then, from the equation of (\ref{vac1c}), we have 
$Z\times a_5=0$.  It  says that 
we may consider the three cases :\ 
1)\ $a_{5\al}=0$, \ 2)\ $Z_\al=0$, \ 3)\ $a_{5\al}\propto Z_\al (\neq 0)$. 
%It turns out that the case 3) includes the case 1) and 2).
% Hence we explain case 3). 

The case 3) solution is given by
%*** vac30 %%%%%%%%%%%%%%%%
\begin{eqnarray}
\vp_\al=\vpbar_\al\{ c_1[\xf]_p+[\xf-l]_p\} \com\q
a_{5\al}=\abar_\al\{c_2[\xf]_p+[\xf-l]_p \} \com\nn
\eta=\mbox{const}\com\q\eta'=\mbox{const}\com\q
\abar_\al=c_3\vpbar_\al\com\q
\vpbar_\al=\frac{g}{2l}\eta^\dag T^\al\eta=
\frac{1}{c_1}\frac{g}{2l}\eta'^\dag T^\al\eta'\com\nn
\chi^3_\al=-g\{ \del(\xf)+c_1\del(\xf-l)\}\eta^\dag T^\al\eta\ ,\ 
Z_\al=d_\al
=-\frac{g}{2l}(1+c_1)\eta^\dag T^\al\eta\ ,\ 
\pl_5(\del A_{5\al})|_{\xf=0,l}=0
\com
\label{vac30}
\end{eqnarray}
%%%%%%%%%%%%%%%%%%%%%%%%%%%%%
where $c_1, c_2$ and $c_3$ are three free parameters.
The meaning of $c_1$ is the scale freedom in the "parallel"
condition of brane sources 
$\eta'^\dag T^\al\eta'\propto \eta^\dag T^\al\eta$
%(the same as that of the free parameter $c$ of Case 1))
,
and that of $c_2$ and $c_3$ is the "free" wave property
of $a_{5\al}$ 
%(the same as that of the free parameters
%$c_1$ and $c_2$ of Case 2))
.
%In this solution, $\vp_\al$ and $a_{5\al}$ appear
%rather equally. 
The bulk {\it scalar}
configuration influences the boundary source fields
through the parameter $c_1$, whereas the bulk {\it vector}
(5th component) does not have such effect. Instead
the latter one satisfies the field equation 
only within the restricted variation (Neumann boundary condition). 
This solution includes the case 1) as  $c_3=0$ and case 2) as $c_1=-1$. 
Another special cases are given by fixing two
parameters, $c_1$ and $c_2$ (keeping the $c_3$-freedom),
as shown in Table 1.

We have solved only (\ref{vac1c}). 
When $m_{\alp\bep}$ and $\la_{\alp\bep\gap}$ are given,
the equations (\ref{vac2a}) should be
furthermore solved for $\eta, \eta', f$ and $f'$
using the obtained result. 
The solutions in the second row ($c_1=-1$) of Table 1 correspond
to the SUSY invariant vacuum, irrespective of
whether the vacuum expectation values of the brane-matter
fields ($\eta$ and $\eta'$) vanish or not. 
For other solutions, however, $d_\al$ depends on
$\eta$ or $\eta'$, hence the SUSY symmetry of the vacuum
is determined by the brane-matter fields. 
The eqs. (\ref{vac2b}) have a 'trivial' solution
$\eta=0, f=0$ (or $\eta'=0, f'=0$) when 
$d_\al=-\frac{g}{2l}\eta^\dag T^\al\eta$ 
(or $d_\al=-\frac{g}{2l}{\eta'}^\dag T^\al\eta'$). 
It corresponds to the SUSY invariant vacuum. 
If the equations have a solution $\eta\neq 0$ (or $\eta'\neq 0$), 
it corresponds to a SUSY-breaking vacuum. 

We see the bulk scalar $\Phi$ is localized on the wall(s) where
the source(s) exists, whereas the extra component of the bulk vector
$A_5$ on the wall(s) where the Neumann boundary condition 
%("free-end" condition ?????) 
is imposed.
The two cases, ($c_1=-1, c_2=-1$) and ($c_1=1/0, c_2=1/0$), 
are treated in [4]%\cite{IM0402}
.

%%%%%%%%%%%%%%%%%%%%%%%%%%%%%%%%%%%%%%%%%%%%%%%%%%%%%%%%%%%%%%%%%%%%%%
%%%%%%%%%%% Stability and Localization               %%%%%%%%%%%%%%%%%
%%%%%%%%%%%%%%%%%%%%%%%%%%%%%%%%%%%%%%%%%%%%%%%%%%%%%%%%%%%%%%%%%%%%%%
{\bf 3}\ {\it Stability}\q
In the present approach, ($\Ncal=1$)SUSY is basically respected. If SUSY is
preserved, the solutions obtained previously
are expected to be stable, because the force between
branes (Casimir force) vanish from the symmetry.
In some cases, we can more strongly confirm the stableness from the topology
(or index) as follows. 
We can regard the extra-space size ($S^1$ radius) $l$ as
an {\it infrared regularization} parameter for
the {\it non-compact} extra-space {\bf R}($-\infty<y<\infty$). 
An interesting case is the $l\ra\infty$ limit of
($c_1=-1, c_2=-1$) in Table 1:\  
%*** higgs2 %%%%%%%%%%%%%%%%
%\begin{eqnarray}
$\vp_\al=-\vpbar_\al l\ep(\xf)\ra 
-\frac{g}{2}\eta^\dag T^\al\eta\eptil(\xf),\ 
a_{5\al}=-\abar_\al l\ep(\xf)\ra 
-c_3\frac{g}{2}\eta^\dag T^\al\eta\eptil(\xf),\ 
\chi^3_\al=-g\del(\xf)\eta^\dag T^\al\eta,\ Z_\al=d_\al=0,\ 
\pl_5(\del A_{5\al})|_{\xf=0}=0$. 
%\pr
%\label{higgs2}
%\end{eqnarray}
%%%%%%%%%%%%%%%%%%%%%%%%%%%%%
%Indeed we can confirm %***CHECK! CHECK!*** 
The above limit is a solution of 
%*** higgs3 %%%%%%%%%%%%%%%%
%\begin{eqnarray}
$S=\int d^4x\int_{-\infty}^{+\infty}dx^5\{\Lcal_{blk}+
{\tilde \del}(x^5)\Lcal_{bnd}\}
\ ,\ 
-\infty <\xf<\infty$, 
%\label{higgs3}
%\end{eqnarray}
%%%%%%%%%%%%%%%%%%%%%%%%%%%%%
where $\Lcal_{blk}$ and $\Lcal_{bnd}$ are the same as before 
except that fields are no more periodic. 
The stableness is clear from the same situation as the kink solution
. On the other hand, in the $l\ra\infty$ limit of ($c_1=1/0, c_2=1/0$)
there remains no localization configuration.

%%%%%%%%%%%%%%%%%%%%%%%%%%%%%%%%%%%%%%%%%%%%%%%%%%%%%%%%%%%%%%%%%%%%%%
%%%%%%%%%%% Conclusion                               %%%%%%%%%%%%%%%%%
%%%%%%%%%%%%%%%%%%%%%%%%%%%%%%%%%%%%%%%%%%%%%%%%%%%%%%%%%%%%%%%%%%%%%%
{\bf 4}\ {\it Conclusion}\q
In the brane system appearing in string/D-brane theory,
the stableness is the most important requirement. 
We find some stable brane configurations in the SUSY
bulk-boundary theory. 
We systematically solve the singular field equation
using a general mathematical result about the free-wave solution
in $S_1/Z_2$-space[1]. 
The two scalars, the extra-component of the bulk-vector
($A_5$) and the bulk-scalar($\Phi$), constitute the solutions. 
Their different roles are clarified. 
The importance of the "parallel" configuration is disclosed. 
The boundary condition (of $A_5$) and the boundary matter fields
are two important elements for making the localized
configuration. 
Among all solutions, the solution ($c_1=-1,c_2=-1$) is expected to be
the thin-wall limit of a kink solution. 

In ref.[4,3]%\cite{IM0402,IM0312}
, the 1-loop effective potential is obtained
for the backgrounds $(c_1=-1, c_2=-1)$. 
In ref.[6]%\cite{IM0403}
, a bulk effect in the 1-loop effective potential
is analyzed in relation to the SUSY breaking. 
We hope the family of present solutions will be used for
further understanding of the bulk-boundary system.

%%%%%%%%%%%%%%%%%%%%%%%%%%%%%%%%%%%%%%%%%%%%%%%%%%%%%%%%%%%%%%
%%%%%%%%%%%%%%%%%%%%%  Acknowledgment %%%%%%%%%%%%%%%%%%%%%%%
%%%%%%%%%%%%%%%%%%%%%%%%%%%%%%%%%%%%%%%%%%%%%%%%%%%%%%%%%%%%%%
%\begin{flushleft}
%{\bf Acknowledgment}
%\end{flushleft}
%The authors thank N. Sakai for valuable comments 
%when this work, still at the primitive stage, 
%was presented at the Chubu Summer School 2002 (Tsumagoi, Gunma, Japan,
%2002.8.30-9.2). A part of
%this work was done when 
%one of the author (S.I.) stayed at
% DAMTP(Univ. of Cambridge,2002.11.22-2003.2.10). He thanks 
%G.W. Gibbons and G. Silva for comments and discussions. 
%The hospitality there is acknowledged. 
%He also thanks the governor of the Shizuoka prefecture for
%the financial support.

%%%%%%%%%%%%%%%%%%%%%%%%%%%%%%%%%%%%%%%%%%%%%%%%%%%%%%%%%%%%%%%%%%%%
%%%%%%%%%%%%%%%%%%%%%       Table 1          %%%%%%%%%%%%%%%%%%%%%%%
%%%%%%%%%%%%%%%%%%%%%%%%%%%%%%%%%%%%%%%%%%%%%%%%%%%%%%%%%%%%%%%%%%%%
$$
%*** mp %%%%%%%%%%%%%%%%
%\begin{eqnarray}
\begin{array}{c|c|c|c}
             &  \begin{array}{c} c_2=-1\\ \pl_5(\del A_{5\al})|_{\xf=0,l}=0\end{array}      
                        & \begin{array}{c} c_2=0\\ \pl_5(\del A_{5\al})|_{\xf=0}=0\end{array}    
                                       &  \begin{array}{c} c_2=1/0\\ \pl_5(\del A_{5\al})|_{\xf=l}=0\end{array} \\
\hline
 & & & \\
\begin{array}{c} c_1=-1 \\ (\eta'^\dag T^\al\eta' \\ \ =-\eta^\dag T^\al\eta)  \end{array}   & 
\begin{array}{c}
\pl_5\vp_\al:\ B\Bbar\\
\pl_5a_{5\al}:\ B\Bbar\\
d_\al=0
\end{array}
          &
\begin{array}{c}
\pl_5\vp_\al:\ B\Bbar\\
\pl_5a_{5\al}:\ B\\
d_\al=0
\end{array}
             &
\begin{array}{c}
\pl_5\vp_\al:\ B\Bbar\\
\pl_5a_{5\al}:\ \Bbar\\
d_\al=0
\end{array}
                                     \\
\hline
% & & & \\
\begin{array}{c} c_1=0 \\ (\eta'^\dag T^\al\eta'=0) \end{array}  &
\begin{array}{c}
\pl_5\vp_\al:\ B\\
\pl_5a_{5\al}:\ B\Bbar\\
d_\al=-\frac{g}{2l}\eta^\dag T^\al\eta
\end{array}
          &
\begin{array}{c}
\pl_5\vp_\al:\ B\\
\pl_5a_{5\al}:\ B\\
d_\al=-\frac{g}{2l}\eta^\dag T^\al\eta
\end{array}
               &
\begin{array}{c}
\pl_5\vp_\al:\ B\\
\pl_5a_{5\al}:\ \Bbar\\
d_\al=-\frac{g}{2l}\eta^\dag T^\al\eta
\end{array}
                                         \\
\hline
 & & & \\
\begin{array}{c} c_1=1/0 \\ (\eta^\dag T^\al\eta=0) \end{array}   &
  \begin{array}{c}
\pl_5\vp_\al:\ \Bbar\\
\pl_5a_{5\al}:\ B\Bbar\\
d_\al=-\frac{g}{2l}\eta'^\dag T^\al\eta'
\end{array}
            & 
\begin{array}{c}
\pl_5\vp_\al:\ \Bbar\\
\pl_5a_{5\al}:\ B\\
d_\al=-\frac{g}{2l}\eta'^\dag T^\al\eta'
\end{array}
                &
\begin{array}{c}
\pl_5\vp_\al:\ \Bbar\\
\pl_5a_{5\al}:\ \Bbar\\
d_\al=-\frac{g}{2l}\eta'^\dag T^\al\eta'
\end{array}
                                            \\
%\multicolumn{4}{c}{\q}                        \\
\multicolumn{4}{c}{
\mbox{
Table\ 1\ \ Various vacuum configurations of
the Mirabelli-Peskin model.
      } 
                  }           \\
\multicolumn{4}{c}{
\mbox{
$B\Bbar$, $\Bbar$ and $B$ correspond to brane-anti-brane, anti-brane
and brane respectively.
      } 
                  }            \\
%\multicolumn{3}{c}{        }\\
\end{array}
%\label{table1 }
%\end{eqnarray}
$$
%%%%%%%%%%%%%%%%%
%%%%%%%%%%%%%%%%  END  of  Table 1 %%%%%%%%%%%%%%%%%%%%%%%%%%%%%

{\bf 5}\ {\it References}\nl
%%%%%%%%%%%%%%%%%%%%%%%%%%%%%%%%%%%%%%%%%%%%%%%%%%%%%%%%%%%%%%%%%%
%%%%%%%%%%%%%%%%%%%%%%%% reference %%%%%%%%%%%%%%%%%%%%%%%%%%%%%%%
%%%%%%%%%%%%%%%%%%%%%%%%%%%%%%%%%%%%%%%%%%%%%%%%%%%%%%%%%%%%%%%%%%
%\begin{thebibliography}{99}
%\bibitem{Rajara82} %*Rajara82*
[1]\ S. Ichinose and A. Murayama, hep-th/0405065, 
Phys.Lett.{\bf 596B}(2004)123\nl
%\bibitem{MP97} %*MP97*
[2]\ E.A.Mirabelli and M.E. Peskin, hep-th/9712214, 
Phys.Rev.{\bf D58}(1998)065002\nl
%\bibitem{IM0312} %*IM0312*
[3]\ S. Ichinose and A. Murayama, hep-th/0401011, US-03-08
"Quantum Dynamics of A Bulk-Boundary System"\nl
%\bibitem{IM0402} %*IM0402*
[4]\ S. Ichinose and A. Murayama, hep-th/0302029, 
Phys.Lett.{\bf 587B}(2004)121\nl
%"Brane-Anti-Brane Solution and SUSY Effective Potential
%in Five Dimensional Mirabelli-Peskin Model"
%\bibitem{Hebec01} %*Hebec01
[5]\ A. Hebecker, \NP{\bf B632}(2002)101\nl
%\bibitem{IM0403} %*IM0403*
[6]\ S. Ichinose and A. Murayama,hep-th/0403080,
Phys.Lett.{\bf 593B}(2004)242. 
%"A Bulk Effect to SUSY Effective Potential in a 5D Super-Yang-Mills Model"
%\end{thebibliography}

\end{document}